\documentclass[aps,prl,showpacs,twocolumn]{revtex4}
\usepackage{graphicx}
\begin{document}
\newcommand {\be}{\begin{equation}}
\newcommand {\ee}{\end{equation}}
\newcommand {\bea}{\begin{eqnarray}}
\newcommand {\eea}{\end{eqnarray}}
\newcommand {\nn}{\nonumber}


\title{  Energy levels and their correlations in quasicrystals  }

\author{ A. Jagannathan and F. Pi\'echon}
\affiliation{Laboratoire de Physique des Solides, CNRS-UMR 8502, Universit\'e
Paris-Sud, 91405 Orsay, France }

\date{\today}

\begin{abstract}
Quasicrystals can be considered, from the point of view of their electronic properties,
as being intermediate between metals and insulators. For example, experiments 
show that quasicrystalline alloys such as
AlCuFe or AlPdMn have conductivities
far smaller than those of the metals that these alloys are
composed from. Wave functions in a quasicrystal are typically
intermediate in character between the extended states of a crystal
and the exponentially localized states in the insulating phase, and this is 
also reflected in the energy spectrum and the density of states. 
In the theoretical
studies we consider in this review, the quasicrystals are described by a pure
hopping tight binding model on simple tilings. We focus on spectral properties,
which we compare with those of other complex systems, in particular,
the Anderson model of a disordered metal. We discuss ``strong" and ``weak" quasicrystals, which
are described by different universal laws. We find similarities and universal behavior, but also 
significant differences between quasiperiodic models and models with disorder.
Like weakly disordered metals, the quasicrystal can be described by the universal level
statistics that can be derived from random matrix theory. 
These level statistics are only one aspect of the energy spectrum, 
whose very large fluctuations can also be described by a level spacing distribution that is
log-normal. 
 An analysis of
spectral rigidity shows that electrons diffuse with a bigger
exponent (super-diffusion) than in a disordered metal. Adding disorder attenuates the singular properties of the
perfect quasicrystal, and leads to improved transport. Spectral properties are also used in
computing conductances of such systems, and to attempt to resolve the experimental enigmas
such as whether quasicrystals are intrinsically conductors, and if so, how conductances depend on the structure.

\end{abstract}
\maketitle

\section{\label{sec:level1}I. Introduction}

\begin{figure}[ht]
\begin{center}
\includegraphics[scale=0.80]{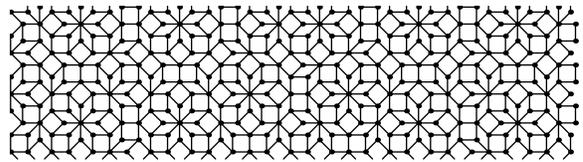}
\vspace{.2cm} \caption {Finite piece of a quasiperiodic structure  }
\label{qcrystal.fig}
\end{center}
\end{figure}

Inside a perfect quasicrystal, such as the Penrose tiling
\cite{penrose} or the octagonal tiling shown in Fig.\ref{qcrystal.fig}, a
freely wandering observer would find a small reoccurring set of
local environments, as in crystals. The sequence of
environments, however, does not repeat periodically, in the case
of the quasicrystal. The repetitivity property states that any given pattern is guaranteed to be
repeated within a distance proportional to the linear size $R$ of the
pattern. Random structures are very different in this respect, in that
the number of patterns increases rapidly with the size of the pattern,
 and in this case identical regions of size $R$ will
be typically separated by
 distances that are
exponentially large for large $R$. This geometric property has implications for the spatial extent
of wavefunctions, as we will see later. Since the repetition is not periodic, one does not have a Bloch theorem in the quasicrystal.
 That the quasiperiodic structure is not random is perhaps
best seen by calculating its structure factor, which turns out to
have sharp Bragg peaks, as in a crystal \cite{levine}. The structure factor of a quasicrystal can also exhibit  
 5-fold 8-fold or 10-fold symmetries that are forbidden for crystals.

A wave packet representing an electron that enters such a
quasiperiodic medium will be scattered by the quasiperiodic
potential, giving rise to a complex superposition of wave vectors
and corresponding phase shifts. What does a typical wavefunction look like, and what
are the allowed energy levels ? Would the electrons conduct an
electrical current, and if so, what factors determine the value of
the conductivity in a real quasicrystal ? Such questions have
been asked, since the discovery by Schechtman et al \cite{shecht} of real quasicrystals in 1984. Some
answers have been provided, by direct experimental measurement,
and in numerical calculations. However, many basic questions
concerning the properties of even the simplest two-dimensional
quasiperiodic structures remain unanswered. 

In this paper we present a review of some electronic properties of a single
 electron in a
quasiperiodic solid in two dimensions. This is
 the lowest dimension in which new
physics due to the interplay between order and disorder is
expected to arise. A well known example is that of disordered
crystalline solids described by the Anderson model
\cite{anderson}, where any disorder, no matter how weak, ``wins
out" over the periodic order and exponentially localizes
wavefunctions, in one dimension. In two dimensions, this tendency
is less marked, and one has the notion of weakly and strongly
localized wavefunctions. A transition between
localized-delocalized transition only occurs, in the Anderson
model, in three dimensions and above. A similar situation could be
expected to hold for the quasiperiodic Hamiltonians. The case of
one dimension is the most tractable, in that analytical methods
can be combined with numerical solutions in order to get a good
appreciation of the expected physics. However, once again, one
dimension is special in that the quasiperiodic potential is almost
always a relevant variable (in the sense of the renormalization
group).

One of the principal symmetries of the quasicrystal is its invariance
under a scale change -- inflations and deflations (to be described in Sec.II). 
The scale invariance of quasiperiodic structures, and in
consequence, the potential seen by the electrons, must play an
important role in determining the spatial dependence and the
energy of the eigenstates. The characteristic singular features
 in the density of states and other electronic
 properties are in fact a result of this symmetry.
There have been attempts to exploit the heirarchical symmetry of
the tiling in order to obtain renormalization equations
\cite{siremoss1} for tight-binding models, but the method breaks down for
the pure hopping models that we discuss here. In a different context, another RG
scheme was introduced for a spin model on this tiling, in order to calculate ground state energy
and local magnetizations
\cite{jag1}.  It is therefore necessary, in the
absence of analytical tools, to take a numerical approach to the electronic problem.
 
 Tilings can be deterministic (``perfect") or
random. It is of interest to study the effects of including
geometric disorder since real quasicrystals may well be
intrinsically disordered in this sense. Random tiling models have
been introduced (see review in \cite{henley}) as the basic templates for
naturally occurring quasicrystals. In contrast to the perfect tiling,
the random tiling has a lower free energy due to the associated
entropy, explaining why materials might choose to
condense with this type of long range order. Careful X-ray diffraction experiments must be
done \cite{boissieu} to distinguish the random tiling from the perfect
tiling, as Bragg peaks occur in the same positions - this shows that the
geometrical constraints are locally strong enough to ensure that
even in random tilings the structure retains long range order with
an infinite correlation length.  

The experimental motivation for our studies comes from a large number of
intriguing results for the
electrical conductivity of quasicrystals. Firstly, the low temperature
electrical conductivity of quasicrystals is anomalously low, many
orders of magnitude smaller than that of its constituent metals,
aluminium, zinc, copper or iron, etc. Secondly, the conductivity rises
with temperature, contrarily to the usual metallic behavior. Thirdly, when
the structural disorder initially present in as-quenched samples
of AlFeCu is removed by annealing, the conductivity gets smaller
\cite{klein}. Similarly, in samples of AlPdMn prepared by
different methods,
 the sample of higher
structural perfection has the lower conductivity \cite{swen,pre}.
Disorder appears to facilitate transport in the quasicrystal,
contrarily to its effect in disordered metals. Finally,
studies of complex alloy system for the AlPdMn with very large
unit cells show that the conductivity was higher for these than
for a single grain quasicrystal of similar composition, although
both alloys had comparable structural quality. This could indicate
that the transport is not just determined by local properties, but also
by the long range quasiperiodic structural order.

Many phenomenological \cite{mayou} and numerical \cite{fuji} studies have
therefore addressed the problem of explaining the magnitude and temperature dependence of the electrical conductivity in realistic models. However studies of even the simplest of tight binding models
show that the properties of the quasicrystal are far more complex than those of crystals. Therefore,
 our focus in this paper is restricted to the simplest of two dimensional quasiperiodic tilings and determining
for these whether states are expected to be localized or extended, whether electrons will conduct,
and how best to describe the complex dynamics in such systems.

\section{\label{sec:level1}I. Definitions and general background}
We consider tight-binding Hamiltonians of the form
\begin{eqnarray}
 H = - \sum_{\langle i,j\rangle} t_i(c^\dag_ic_j + c^\dag_jc_i) + \sum V_i c_i^\dagger c_i
\label{tbham}
\end{eqnarray}
in terms of onsite electron creation and annihilation operators
$c_i^\dag$ and $c_i$ on a variety of two dimensional structures.
 The
first term allows for hopping between connected sites with
amplitudes $\{t_i\}$, while the second term allows for variations in
the onsite potential. Although this Hamiltonian is grossly
oversimplified, it contains the essential information about the
quasiperiodic structure, namely, its geometry. For a discussion of
the tight-binding Hamiltonian and its properties, particularly for
one-dimensional quasicrystals, the reader is referred to
\cite{sire}. In two dimensions, there have been many calculations
for this type of model and its dual version on the
octagonal (Ammann-Beenker) and the Penrose tilings
\cite{oda,koh,suth2,kumar,toki,benza1,benza2,jag2,pj,zhong,zijl,tsune,tsune2}.

We will now simplify the Hamiltonian even further by setting
henceforth all of the hopping amplitudes equal to a constant,
$t=1$, and all the $V_i =0$. This is because we wish to focus
exclusively on the effects of the quasiperiodic connectivity
between atoms. Any variations in the hopping amplitude, or of the
local potentials has the effect of accentuating the quasiperiodic
modulation and opening gaps in the spectrum.
For example, if one includes onsite potentials $V(\bf{r}_i)$
that are proportional to the site coordination number, one gets, even for
moderate values of $V$, bands separated by true gaps, and
wavefunctions which have a quasi one-dimensional character
\cite{benza1,zijlstra}.

The Anderson model for disordered conductors belongs in this class
of tight-binding Hamiltonians. The disorder is realized by taking random
hopping amplitudes and/or random onsite potentials or both. Such
models are used, for example, to study critical properties at the
metal-insulator transition in three dimensions. Spectral
statistics in the Anderson model have also been discussed
intensively, and compared with those of other classically chaotic
systems. These studies ( see for example the review in
\cite{andersonrefs}), along with the predictions of random
matrix theory provide the background for the present discussion of
the quasicrystal.

\subsection{Obtaining planar quasiperiodic structures}

\begin{figure}[ht]
\begin{center}
\includegraphics[scale=0.50]{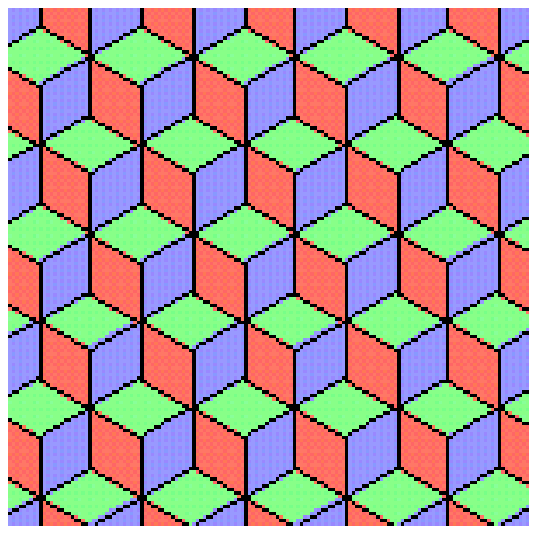} \hskip 2cm
\includegraphics[scale=0.50]{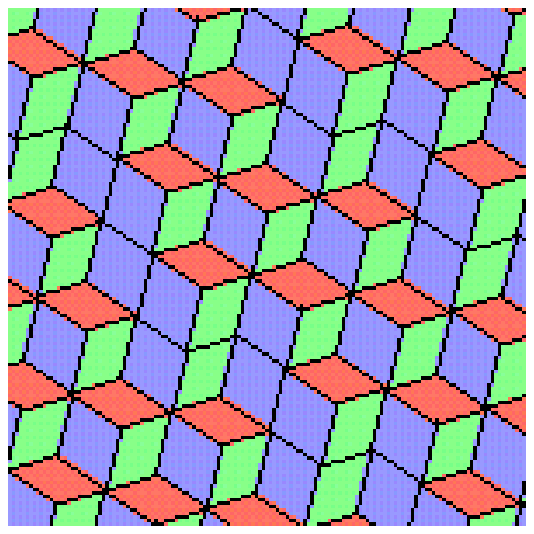}
\vspace{.2cm} \caption{a)a periodic projection and b) an aperiodic projection of the cubic lattice}
\label{dice.fig}
\end{center}
\end{figure}

A standard method of obtaining quasicrystals and their
approximants consists of projecting down from a higher dimensional
cubic lattice \cite{levine}. A
subset of points belonging to an infinite strip of this hypercubic
lattice is projected onto the physical plane. Fig.\ref{dice.fig}
(taken from \cite{durandref}) shows the projection
onto the plane of a portion of the simple cubic lattice when the
projection is rational (left-hand figure), and for an irrational
orientation (right figure) where the structure never repeats. In
these figures, the different faces of the hypercube are projected
onto tiles of different colors. A similar approach can be used to
obtain the Penrose (projection from five dimensions), the
octagonal tiling (projection from four dimensions) and the weakly quasiperiodic tiling 
described in
Sec. III (projection of the simple cubic lattice). 

In calculations, the structures considered are often finite
periodic samples, called periodic approximants, obtained by
rational cuts in the hyperspace. The unit cell of N sites can be made arbitrarily large,
by modifying the plane of projection. The use of approximants has the
advantage of eliminating surfaces and thus spurious energy levels
-- at the price of introducing a few defects in the structure.
However, these are of diminishing importance as the approximant
size is increased, and the true quasicrystal structure is
approached. For the square approximants of the octagonal tiling, we
 assume boundary conditions $\psi(x+L,y) = e^{ik_xL} \psi(x,y)$;
$\psi(x,y+L) = e^{ik_yL} \psi(x,y)$ where $k_\mu = 2\pi n_{\mu}/L$ and $L$ is the 
repeat length of the approximant in each direction. An arbitrary
external magnetic flux through a gauge field $\phi$, can also be
introduced by suitably modifying the hopping amplitudes by a phase
factor. As we will see, the addition of magnetic flux changes the
symmetry properties of the Hamiltonian, with a corresponding
change in spectral characteristics.

The $N\times N$ real symmetric matrix representing $H$ is numerically diagonalized for each $\{k_x,k_y\}$ and the set of eigenvalues is used to compute the density of states and interesting statistical properties of energy levels. For purposes of computing the density of states, assuming large enough systems, it is in fact sufficient to compute the energy spectrum at $\vec{k}=0$. The two main statistical quantities considered are the nearest neighbor level spacing distribution $P(s)$ and the level rigidity function $\Sigma^2(E)$. These quantities are introduced now, along with some known results and their interpretation.

\subsection{ Level spacing statistics}

After ordering the energy levels corresponding to a given
$\vec{k}$ and parity, we can compute the set of spacings $s_n =
E_{n+1} - E_n$. The mean level spacing is denoted by $\Delta$.

If
the density of states $\rho(E)=N^{-1}\sum_n \delta(E - E_n)$ were
constant as a function of the energy, one would simply go on to
construct the histogram representing the probability distribution
$P(s)$ defined by Eq.\ref{ps2}.
However, in general, the density of states $does$ depend on the
energy. In this case, the spacings must be redefined or
``unfolded" so as to eliminate this trivial dependence on the
position of the levels in the band.  A systematic way to redefine
spacings consists of fitting the integrated density of states by a smoothed
``classical" part. This smoothed curve is then used to
define a set of unfolded spacings
(see \cite{boh}). $P({s})$ is now
determined from the set of unfolded spacings $\{s_n\}$ using

\bea P({s}) = \frac{1}{N} \sum_{n=1}^N \delta({s}-{s}_n) \eea

The issue of the statistical analysis of energy levels was
raised first for the spectra of large nuclei. The first
theoretical calculations of level spacing distribution in complex
systems were carried out by Wigner and by Dyson, for random
matrices having independently distributed gaussian random matrix
elements. The result depends on the class of matrix \cite{mehta}.
Three cases were distinguished: real symmetric matrices (gaussian
orthogonal ensemble or GOE), complex unitary matrices
(gaussian unitary ensemble or GUE) and, finally, symplectic random
matrices (GSE)

\bea P_{WD}(s) = A_\beta s^\beta{e}^{-C_\beta s^2} \label{wd} \eea

where $\beta =1,2,4$ for GOE , GUE and GSE respectively. Values
for the constants are, in the GOE case, $A_1 = \pi/2$ and
$C_1=\pi/4$. These distributions are written for spacings that
have been normalized so that the mean level spacing $\Delta =1$.
An impressive array of complex systems turn out to conform to one
of these three expressions of $P_{WD}(s)$: heavy nuclei, chaotic
billiards, correlated fermions in regular crystals, etc. The
Wigner-Dyson distributions thus appear to be a universal feature
of classically non-integrable systems where the motion is ergodic.

For the Anderson model, in the metallic regime many studies have
confirmed that the level spacings obey GOE statistics. If time
reversal symmetry is broken, by adding an external magnetic field,
or by including a magnetic flux then GUE statistics are found. The
third class of problems with $\beta=4$ has also been realized in
metals containing spin-orbit scatterers.

\begin{figure}[ht]
\begin{center}
\includegraphics[scale=0.80]{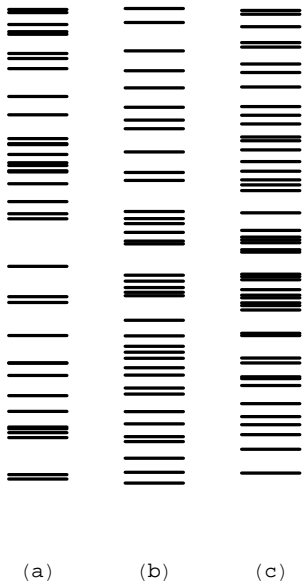}
\vspace{.2cm} \caption {Examples of spectra a) Uncorrelated energy
levels b) disordered metal energy levels c) 
octagonal tiling energy levels} \label{spectra.fig}
\end{center}
\end{figure}

 A feature of note in the level spacing distributions
$P_{WD}$ is the fact that they tend to 0 as $s\rightarrow 0$. This
expresses the unlikelihood of levels being degenerate in this
system - the phenomenon of so-called {\it level repulsion}. Level
repulsion occurs when electron wavefunctions are extended in the
available volume, and when there is, in consequence, a
non-negligible spatial overlap between states. In systems where
wavefunctions are strongly localized, on the contrary, states in
different regions of the sample do not overlap, and spatial
correlations are negligible, there is no bar to two different
states having the same energy. This is the reason that a discrete
level spectrum corresponding to an insulator can be distinguished
visually from that of of a weakly disordered metal. In
Fig.\ref{spectra.fig} three different kinds of spectra
are illustrated (energies have been scaled so
that there is the same mean spacing between levels for all three examples).
In the insulator (left hand figure), distances
between the energy levels of the insulator fluctuate greatly, and
levels can approach arbitrarily close to each other. The middle figure represents energy levels of
a disordered metal and was obtained by diagonalizing the Anderson model Hamiltonian. Here, 
levels repel each other due to non-negligible wavefunction overlap, 
and so there are fewer very small or very large spacings. Finally, the quasicrystal
spectrum is intermediate between the first two examples (right hand figure).

Other probability distributions can be found in the literature,
corresponding to other physically interesting problems. In cases
where the classical dynamics of the electron is integrable, such
as the motion of electrons in a crystal, the probability
distribution of the set of all spacings is governed by a different
law, the Poisson distribution,

\bea P_P(s) = {e}^{- s} \label{ps2} \eea

This behavior is found, as the name indicates, whenever the
levels are randomly distributed according to a Poisson
probability (the left hand figure in Fig.\ref{spectra.fig}).
It is the level spacing distribution in the strong
localization regime of the Anderson model. At the critical point
of the metal-insulator transition, there have been proposals
\cite{shklov} for
 another form of $P(s)$ which is a hybrid between the
 exponential dependence at large $s$ of $P_P$ and the small $s$
 level repulsion of $P_{WD}$:
 written above:

\bea P_c(s) = A_\beta s^\beta  \exp ({-as^{(2-\gamma)}} \eea

where $\beta$ has the values 1 or 2 or 4 as mentioned earlier.

The semi-Poisson distribution $P_{sp}$ is a special case of this
type of law, with $\beta=\gamma=1$. The name semi-Poisson comes
from the fact that this is the distribution for the distances
between the {\it midpoints between levels} given a set of randomly
chosen energy levels. There is level repulsion, but then the
probability rises to a maximum and decays exponentially, as in the
Poisson distribution. At the critical point, wavefunctions are
expected to have multifractal character. This type of wavefunction
leads to level repulsion, unlike states in the localized regime,
but the effect is not as strong as that in the metallic regime.and
Such multicritical wavefunctions have been linked to the level
statistics at the critical point of the Anderson model
\cite{chalker}. This type of law has been observed to hold in the
Anderson model if one averages the distribution over all boundary
conditions \cite{kravt,evan1,braun,batsc}, at the critical point
of the Harper model \cite{evan2} and in a disordered two particle
Hubbard model \cite{waintal}. Another example with such statistics
are the pseudointegrable billiards, 
studied by Bogomolny et al
\cite{bogo}.

\subsection{The spectral rigidity function}
As we mentioned in the last section, the system of energy levels
displays a certain amount of rigidity. One measure of this
rigidity is given by the variance of the number of levels in an
energy interval of width $E$:

\bea \Sigma^2(E) = \langle (N(E)-\langle N(E)\rangle)^2\rangle
\eea

where $N(E)$ is the total number of levels within an interval of
width $E$.
The average is taken over all positions of $E$ in the spectrum.
 In random matrix
theory, the rigidity has been calculated to be a relatively slowly
increasing function of $E$, with

\bea \Sigma^2_{RMT}(E) = \frac{1}{\beta \pi^2} \ln(2\pi E/\Delta)
\eea

In comparison, the rigidity of a completely random set of levels,
where the $P(s)$ is Poissonian, one has a linear dependence

\bea \Sigma^2_P(E) = E \eea

At the critical point, the spectral rigidities are also linear for
small $E$, but with a smaller slope $\Sigma^2 \sim \chi E$ ($\chi <1$), since correlations between
levels begin to kick in, and lead to level repulsion. The
semi-Poisson form will be described in more detail in the third
section, for the case of the weakly quasiperiodic tiling.

 The spectral rigidity function can be obtained from the
two-level correlation function $K(E')$ defined by $K(E')=
\langle N(E)N(E+E')\rangle/\langle N(E)\rangle^2 - 1$. One can
check that $\Sigma^2(E) = 2 \int_0^E dE' (E-E')K(E')$. Argaman et
al \cite{arga} established a connection between the dynamics of an
electron moving in the medium and the spectral form factor
 $K$. Adapting their calculations to a situation where
an electron diffuses with an exponent $\sigma$ defined in terms of
the root mean square distance $d$ explored in a time $t$ by $d
\sim t^\sigma$, then the probability of return to the origin
scales as $ p(t) \sim t^{1-d\sigma}$ where $d$ is the spatial
dimension. It can be shown (see for example \cite{akkmont}) that
$p(t)$ is proportional to $K(t)/t$ where $K(t) = \frac{1}{2\pi}
\int K(\omega) e^{-i\omega t} d\omega$. Transforming back
to energy variables the power law in time for $K$ becomes a power
law in the energy for $\Sigma^2$,

\bea \Sigma^2(E) \sim E^{d\sigma} \label{sigma2.eq} \eea

Thus, if the rigidity function does indeed follow such a power
law, one can from it deduce the exponent $\sigma$ describing
quantum diffusion in the medium. For disordered metals, one has
normal diffusion, $\sigma=\frac{1}{2}$, whereas as we will see,
the motion in the quasicrystal is superdiffusive and $\sigma
>\frac{1}{2}$.

In numerical calculations on disordered metals one usually finds
different regimes of behavior of the spectral rigidity depending
on the energy. The crossover energy scale between low and high
energies is given by the Thouless energy, $E_{Th} = \hbar/t_L$,
where $t_L$ is the time taken for the electron to diffuse to the
boundary of the sample. This time depends on the diffusion
exponent via $t_L \sim L^{1/\sigma}$. For small energies $E <<
E_{Th}$, the behavior of $\Sigma^2$ is governed by the long time
dynamics, which is completely ergodic since the electron has
reached the boundary a large number of times. At such energies,
the RMT logarithmic dependence is thus expected. For $E >>
E_{Th}$, however, one expects to find a power law dependence,
corresponding to the diffusive electron motion at times shorter
than $t_L$.

\subsection{Wavefunctions and quantum diffusion}

The wavefunctions of the tight-binding Hamiltonian
\ref{tbham} are solutions of $H \psi^{(n)} = E_n\psi^{(n)}$. The set of
amplitudes $\psi_i^{(n)}$ ($i=1,N$, where N is the number of
sites) is numerically evaluated for each of the energy eigenvalues
$E_n$. To determine whether a wavefunction is localized or
extended, one usually computes the inverse participation
ratio (IPR) defined by

\bea P^{-1} = \sum_{i=1}^N \vert \psi_i^{(n)}\vert^4 \eea

namely, the second moment of the probability
density. For a localized state, the IPR remains constant as $N$
increases, since such a state involves a finite fixed number
of ``participating " sites. For the extended metallic state, where
a finite fraction of the total number of sites participate, the
IPR decreases as the inverse of $N$. In quasicrystals, an
intermediate situation obtains, with IPR $\propto N^{-\beta}$
where $\beta$ is close to but less than 1 \cite{grimm}.

 The IPR is in fact a very limited probe of the real space
structure of the wavefunction in the quasicrystal. A typical
wavefunction has an average behavior that decays slowly, but also
has extremely large fluctuations from one site to the next as the
figure shows. If one calculates
the exponents that describe the scaling of the moments of
$\psi(E)$ with $N$, one finds that the exponents are not just
multiples of each other -- this is the property of
multifractality. A complete description of such ``critical"
wavefunctions can only be given by specifying the complete set of
moments of such a wavefunction. Such wavefunctions have been
studied in some detail for one dimensional systems like the
Fibonacci chain. They have also been studied at the critical point
of the Anderson model and the Harper model \cite{evan3,evan2}.

The diffusion of a wave packet on such tilings can be studied by
computing the time autocorrelation function $C_i(t)$, where

\bea C_i(t) = t^{-1} \int_0^t  \vert \psi_i(t')\vert^2 dt'
\eea

which gives the probability of the electron staying at the initial site i as a function of time. The mean square displacement for an electron that was initially located at the site $i$ is

\bea r^2(t) = \sum_j \vert r_j - r_{i}\vert^2 \vert
\psi_j(t)\vert^2 \eea

Again, one can determine exponents describing the long time
behavior of these quantities, $C_i(t) \propto t^{-\delta}$ and
$r(t) \propto t^\sigma$. It is this exponent $\sigma$ that has
been related (see previous subsection) to the power law behavior
of the spectral rigidity function $\Sigma^2(E)$. Contrarily to the
case of random systems where self-averaging leads to the same
exponent regardless of which site is chosen as the origin, in the
quasicrystal, $\sigma$ and $\delta$ depend not only on the wave
packet energy but also the site chosen as the origin.

\subsection{Transport}

Thouless \cite{thounum} proposed a measure of intrinsic conductance
that is based on the sensitivity of each level to changes of the
boundary condition. The energy of a localized state is not
affected by a change of boundary condition, whereas the energy of
an extended state is. Thus if a level shifts as a function of
$\vec{k}$, the magnitude of the shift is an indication of the
capacity for transport of that state. One can specify a ``band
width" for each band by $W_n \sim E_n(0,0) - E_n(\pi,0)$. The
Thouless number is given by

\bea g_{th}(E_n) = \frac{W_n}{E_{n+1}-E_{n-1}} \eea

Another quantity that has been used as a quantitative measure of
the transport property of random systems is the set of energy
level curvatures $c_n = \partial^2 E(k)/\partial k^2
\vert_{k\rightarrow 0}$.

A theoretical expression for the distribution of $c$ (parametric level statistics) in the case
of weakly disordered metals has been obtained in RMT (see \cite{basu}):

\bea P_\beta(c) = \frac{a_\beta}{(b_\beta + c^2)^{(2+\beta)/2}}
\eea where $\beta=1,2$ or 4. This law has been verified for
disordered systems \cite{which}.

In the case of the quasiperiodic tilings, both the Thoulesss
number and the curvatures have huge fluctuations, as we will see
below. The form of the distribution does not agree with the RMT
form.

\section{II. A case of strong quasiperiodicity: the octagonal tiling}
In this section we will take up the analysis of the octagonal
tiling with and without disorder. The word ``strong" distinguishes
this tiling from the weakly quasiperiodic tiling, the GRT,
discussed in the next section. The octagonal tiling has an
eight-fold symmetric diffraction pattern, compared to another
strongly quasiperiodic tiling - the Penrose tiling, which has
ten-fold symmetry. This is not an exact rotational symmetry in
real space as in ordinary crystals, instead, it reflects the fact
that any finite domain of the octagonal(Penrose) tiling, occurs
with equal probability in each of eight (respective five) possible
orientations.

\subsection{ Samples with and without disorder}
\begin{figure}[ht]
\begin{center}
\includegraphics[scale=0.45]{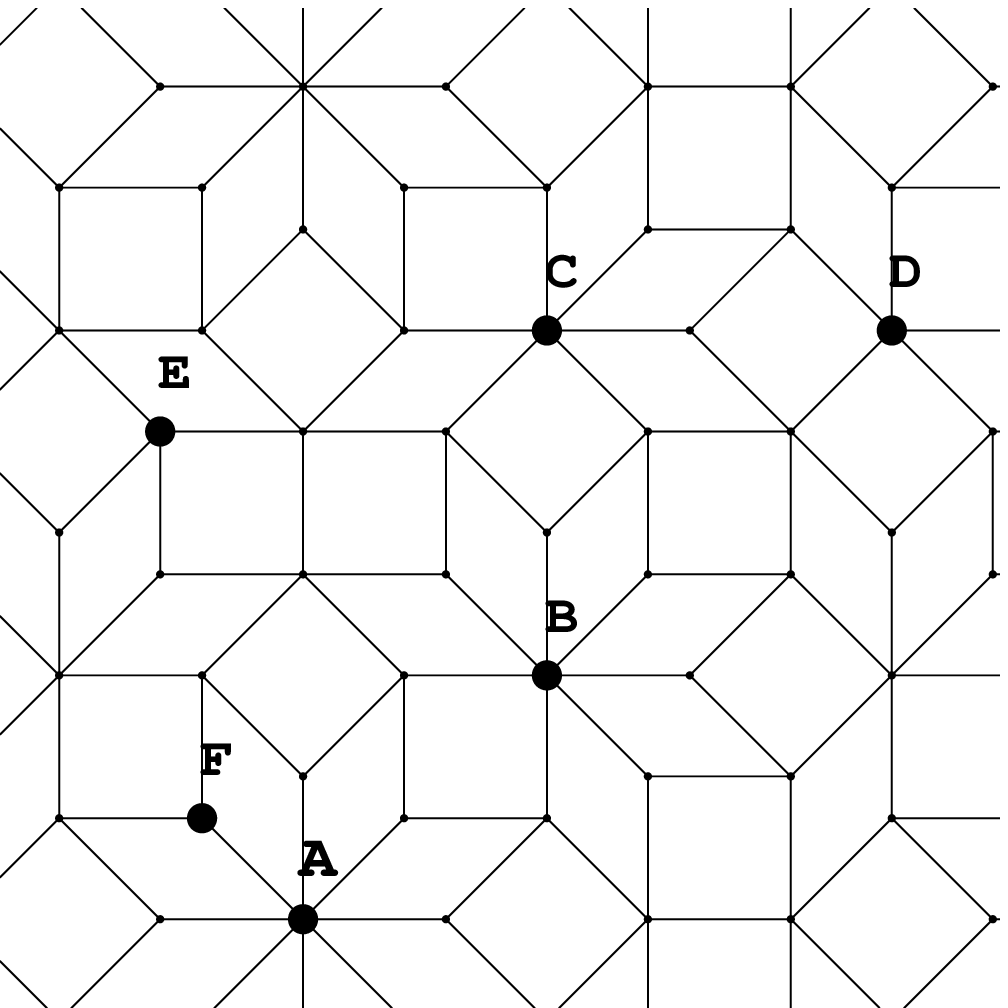}

\includegraphics[scale=0.55]{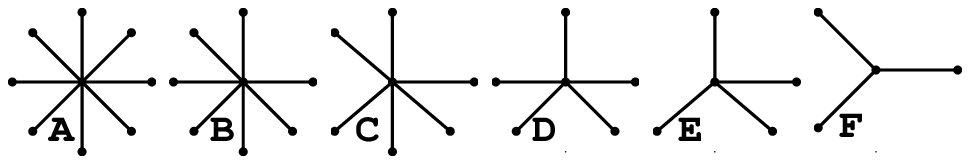}
\vspace{.2cm} \caption{a) A square approximant of the perfect octagonal tiling b) The six local environments }
\label{octag.fig}
\end{center}
\end{figure}

Fig.\ref{octag.fig}a) shows a perfect (disorder-free) square
approximant and the six types of vertices present in it. The
infinite tiling and its approximants are composed from squares and
$45^\circ$ rhombuses. The infinite octagonal tiling can be
obtained by projection from a higher dimensional cubic structure.
For numerical studies, it is preferable to work with samples that
do not have a boundary, the periodic approximants of the octagonal
tiling \cite{oguey}.

Some details about the properties of the octagonal tiling may help
in understanding the structure better. For more details, the
reader is referred to \cite{soco,duneau,baake}. The infinite
tiling and its approximants has six local environments as shown in
fig.\ref{octag.fig}. These are labelled A,B,...,F and
correspond to coordination numbers $z=8,7,6,...,3$ respectively.
It is easy to show using the cut-and-project scheme, for example,
that the relative frequencies of occurrence of each type of vertex
are

\begin{eqnarray}
 f_A=\lambda^{-4};
 f_B=\lambda^{-5}; f_C=2\lambda^{-4}; \nonumber \\
f_{D1}=\lambda^{-3}=f_{D2} ; f_E = 2\lambda^{-2}; f_F = \lambda^{-2}
\nonumber
\end{eqnarray}

\begin{figure}[ht]
\begin{center}
\includegraphics[scale=0.40]{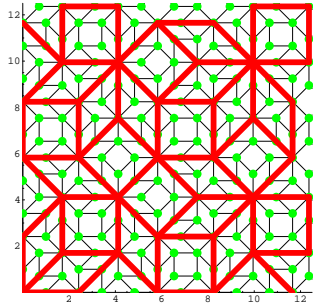}
\vspace{.2cm} \caption{An approximant and its inflated version, showing the old and new tiles}
\label{inflat.fig}
\end{center}
\end{figure}

where $\lambda=1+\sqrt{2}$. This irrational is fixed by the
condition of eight-fold symmetry of the quasicrystal and its
equivalent on the Penrose tiling is the golden mean $\tau
=(\sqrt{5}+1)/2$. An important symmetry of the quasicrystal is
its self-similarity under a discrete scale transformation called
inflation(deflation). Fig.\ref{inflat.fig} shows how to redefine
the tiles or to decimate sites, so as to
obtain an inflated version of the tiling. The density of sites on
the infinite tiling is reduced by a factor $\lambda^2$ after an
inflation. The rational approximants transform into each other
under successive operations of inflation or deflation.

 In
\cite{pj}, we have diagonalized the tight-binding Hamiltonian
of Eq.\ref{tbham} for increasing system sizes, $N_s =
239,1393,8119$, using an extended Lanczsos routine that yields all
distinct energy eigenvalues.

\begin{figure}[ht]
\begin{center}
\includegraphics[scale=0.30]{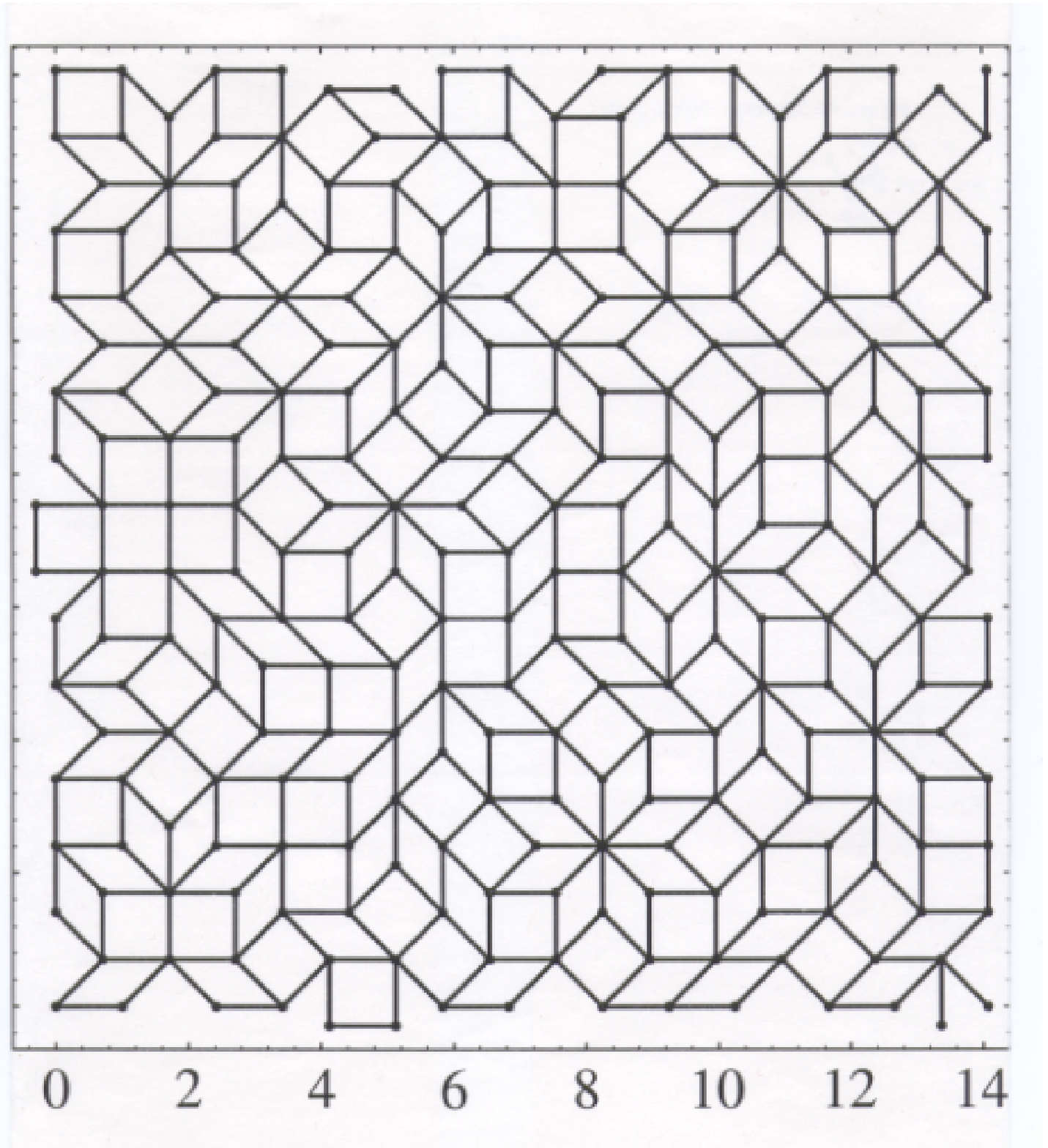} \\

\includegraphics[scale=0.30]{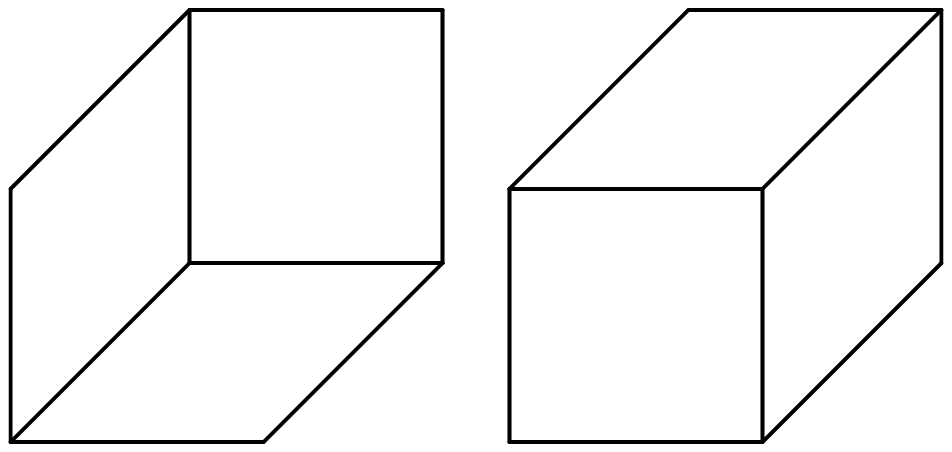}
\vspace{.2cm} \caption{a)a square sample of the randomized octagonal tiling \\
b) a tile flip (or phason) operation}
\label{rando.fig}
\end{center}
\end{figure}

Fig.\ref{rando.fig} shows a sample obtained after one performs a
certain number of operations called ``phason flips"  wherein one
permutes the positions of a square and two rhombuses in the
interior of a hexagonal shaped region. Iterating this operation at
randomly chosen locations on the square approximants, one obtains
a structure that has new environments that are not present in the
original tiling. (Note: in the two dimensional case, contrarily to three
dimensions, such randomizing leads to a reduction of the
peak intensities in the structure factor \cite{henley}).

\subsection{\label{sec:level2} Density of states of the octagonal tiling}

\begin{figure}
\input{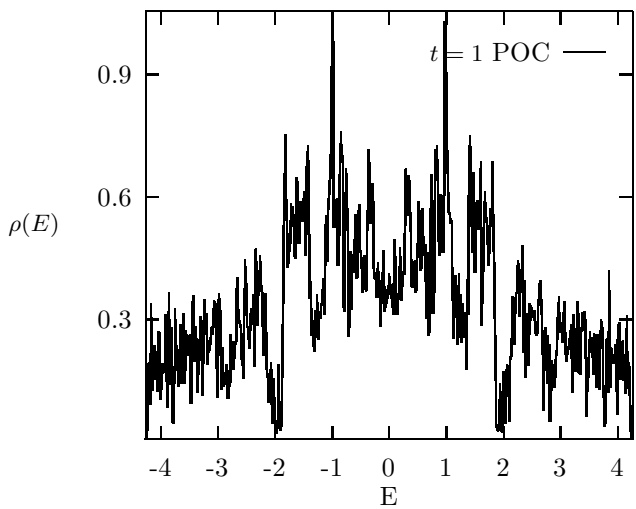}
\caption{DOS of a perfect sample (from \cite{piechonthese})}\label{dosperf.fig}
\end{figure}

\begin{figure}
\input{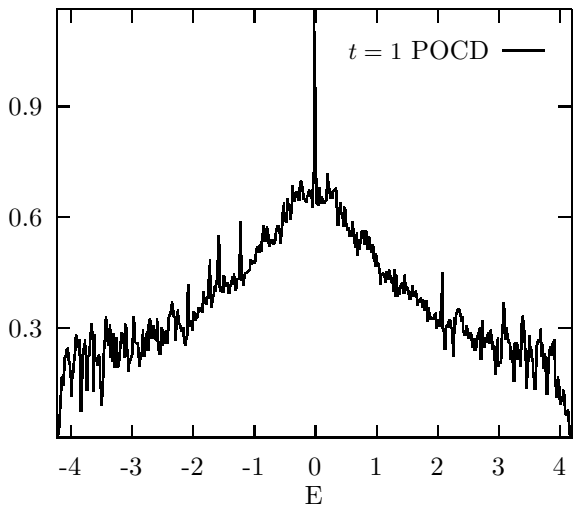}
\caption{DOS of a randomized sample (from \cite{piechonthese})}\label{dosrand.fig}
\end{figure}

Typical histograms representing the density of states (DOS) for
the randomized and for the perfect octagonal tiling are shown in
the figures \ref{dosperf.fig} and \ref{dosrand.fig}.
(In fig.\ref{dosperf.fig} a delta function at $E=0$
corresponding to localized ring-like states around the eight-fold
symmetric sites has been subtracted). Both graphs show a jagged
dependence on the energy, with bigger fluctuations in the perfect
case. The difference between the two DOS is not merely
quantitative, but qualitative. The fluctuations in the randomized
sample are bounded, and do not give rise to pseudogaps, where the
DOS plunges to zero, as in the perfect system. When the DOS
of samples of different sizes are plotted, it becomes evident that
there is no smooth behavior of the DOS in the perfect case:
fluctuations are strong as $N$ is varied at a fixed energy, and
vice versa. This behavior is investigated in more detail below,
where we consider the energy spacings for the two types of
systems. A multifractal analysis  \cite{pj} gives a fractal dimension of the DOS equal to
one, for the perfect tiling, which thus has a single band.

\subsection{The level spacings distribution function $P(s)$ of
the octagonal tiling}

In analysing spacings between energy levels, one has first to
classify the levels in groups according to their symmetry
property. In the case of the approximants, we have one exact
symmetry, namely that by reflection with respect to one of the
diagonals of the square, $x=y$. There are other symmetries - the
infinite tiling has a bipartite structure and so the spectrum is
an even function of the energy $E$. This is broken by the
approximants, but can be restored by, for example, quadrupling the
unit cell so that the boundary conditions do not frustrate the
wavefunctions. In addition, the approximants have an approximate
rotational symmetry \cite{grimm}. In the calculations mentioned
below, levels were grouped according to even(odd) parity under
reflection, but not separated into any further subgroups.

\begin{figure}[ht]
\begin{center}
\includegraphics[scale=0.60]{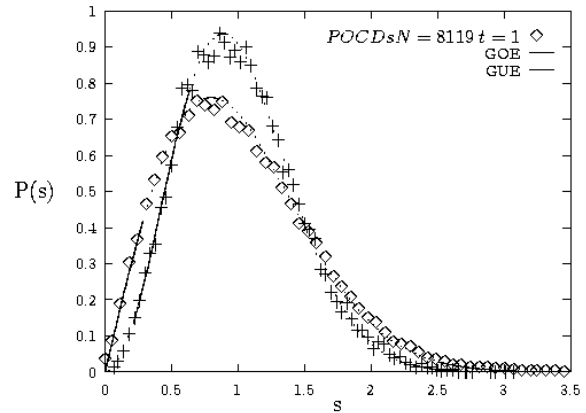}
\vspace{.2cm} \caption{P(s) for the randomized octagonal tiling
with and without inclusion of a magnetic flux (from \cite{piechonthese})}
\label{psrand.fig}
\end{center}
\end{figure}

In most situations, it is possible to determine the unfolded
spacings without any ambiguity by separating the underlying smooth
energy dependence of the DOS. In the case of the randomized
tiling, for example, we are able to define a locally smooth DOS
and use this to renormalize spacings. The distribution obtained in \cite{pj} for these
renormalized spacings is shown in Fig.\ref{psrand.fig}, for two
boundary conditions corresponding to zero and nonzero magnetic
flux respectively (as explained further below). The data points
are shown along with the Dyson-Wigner GOE probability distribution
functions corresponding to $\beta=1$ and 2, as expected.

We found, furthermore, that the same GOE distribution is obtained
when one introduces energetic (diagonal) disorder on the perfect
octagonal tiling \`a la Anderson.

\begin{figure}[ht]
\begin{center}
\includegraphics[scale=0.70]{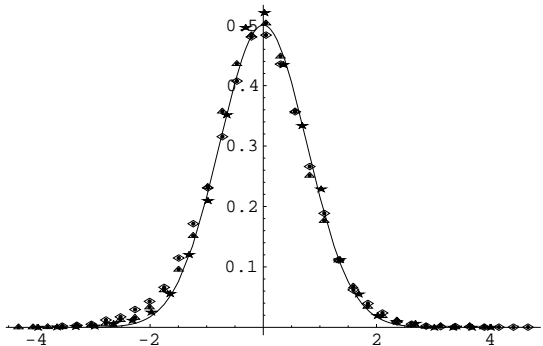}
\vspace{.2cm} \caption{ $P(s)$ for the perfect tiling for bare
spacings for $N$=239 (stars),1393 (triangles) and 8119(diamonds) (curves were shifted to that
the peak positions coincide)} \label{lognorm.fig}
\end{center}
\end{figure}

Moving on to the case of the perfect tiling, the density of states
fluctuations occur on all energy scales, so that it is not obvious
how to separate out a ``classical" part. The same type of problem
occurs at the critical point of the Anderson and the Harper
models. There, it is possible to perform an averaging over all
possible boundary conditions, and determine a local value of the
DOS. This was used to unfold the spacings and thus obtain the
critical level statistics. In the case of the quasicrystal, we
first calculated the distribution of the bare spacings, i.e.
without unfolding. The bare spacings turn out to follow quite well
a log-normal distribution \cite{pj}

\bea P(s) = \frac{1}{2 \pi sB} {\rm e}^{-(\ln s - \ln
s_0)^2/{2B}} \label{lognorm} \eea

where the peak position, $\ln s_0$, shifts to the left as
the size of the tiling is increased. This is, of course, expected
as the mean level spacing decreases $\Delta \approx W/N$ where $W$
is the band width. Fig.\ref{lognorm.fig} shows the data for three
system sizes, after the distributions were shifted so that the
peaks coincide. Note that, unexpectedly, the width of the
distribution, $B$, does not depend on the system size. An
explanation is given in the section below where we outline the
relation between the bare spacing and the unfolded spacing
distributions. There are deviations from the log-normal
curve: for $s$ small the distribution found is linear, and for large $s$
one finds a fit to a power law \cite{piechonthese}.

The log normal $P(s)$ obtained for the bare spacings corroborates
a fact already noted, namely the density of states fluctuates
enormously in the perfectly ordered system. Next we consider
possible ways of unfolding the spacings. The
standard unfolding technique, involves a smoothing of the
integrated DOS as described in \cite{boh}. This smoothing has to
be carried out on an energy scale that is small compared to the
density of states fluctuations but large compared to the average
nearest neighbor spacing, a condition that is not possible to
respect, strictly speaking, in a quasicrystal. In an alternative
definition, we divided the spacings by the local average
value of the spacing defined over M consecutive levels: $\langle
s_i\rangle = (E_{i+M} - E_{i-M})/2M$. When M is large, then the
renormalizations
 depend only weakly on the energy. In this case, the renormalized
 spacings obey a distribution close to that of the bare spacings.
 At the other extreme, for $M=1$, the renormalization factor fluctuates strongly from
 one level to the next.
 In this case the renormalized spacings will obey a very different
probability distribution -- this turns out to be the ubiquitous
Wigner-Dyson distribution \cite{piechonthese}. As figure
\ref{fred.fig} shows,
 one goes continuously from the log normal distribution to the Wigner-Dyson form simply
 upon changing the value of $M$ in the unfolding procedure.
 This was also confirmed by Zhong et al \cite{zhong} using the standard
unfolding procedure given in \cite{boh}. A calculation of the
level spacing distribution for a large but finite patch of the
octagonal tiling confirmed that the same distribution is obtained
in that case as well, despite the open boundary conditions
\cite{grimm}.

The relation between the two types of level spacing distribution
is explained in terms of the heirarchical structure of the
quasicrystal in the next subsection.

\begin{figure}[ht]
\begin{center}
\includegraphics[scale=0.60]{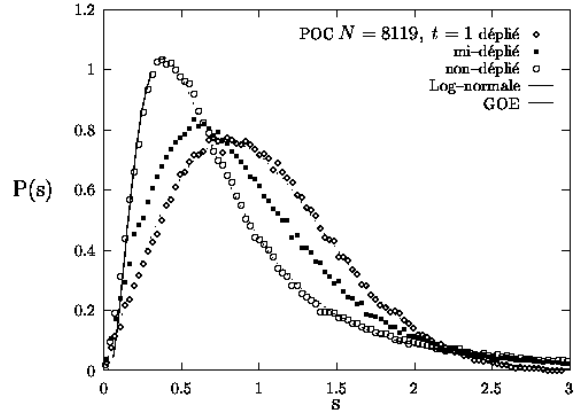}
\vspace{.2cm} \caption{Statistics of levels for three cases: no
level-unfolding, partial unfolding and maximal unfolding 
(from \cite{piechonthese})} \label{fred.fig}
\end{center}
\end{figure}

 \subsection{Relation between folded and unfolded statistics}
As the previous section showed, there does not appear to be a single way to define
unfolded spacings for the infinite tiling, as the notion of ``smoothing" is not
well defined for this situation. In \cite{jagconv} we have proposed a way to overcome this difficulty, which also resolves the issue of the relation between the two statistical distributions $P(s)$ and $P(\tilde{s})$.
When considering level statistics of
 approximants, one can consider using the DOS of a smaller
 approximant to renormalize spacings of the next approximant in the series.
 Recall that approximants are related to each other by inflation/deflation
 transformations. In this way one can write the probability
 distribution of the bigger system in terms of a convolution
 involving the distribution function of the smaller approximant.
 Iterating the process, and taking the initial distribution
 to be of Wigner-Dyson form, one can show that after a few iterations, the distribution
 for the spacings approaches the log normal form. In \cite{jagconv} we have analytically calculated
the variance of the log normal distribution obtained in the large size limit,
and we showed in particular, that it is
 independent of the system size, in agreement with the numerical
 calculations described earlier.

\subsection{The effect of adding magnetic flux}
The time reversal symmetry of the Hamiltonian in eq.\ref{tbham}
can be broken by adding an external field, or, more simply, by
imposing a flux via boundary conditions. This can be done by
identifying two of the edges of the approximant (similarly to when one wraps a graphene sheet to form a nanotube)
and considering a flux tube along the axis of the resulting cylinder. 
A magnetic flux $\phi$ along the axis of the cylinder will
then create a total phase shift of $e^{i \pi\phi}$ on a closed
loop around the flux tube. This phase shift is easily implemented by
taking $\psi(x+L,y)= e^{i \pi\phi} \psi(x,y)$
$\psi(x,y+L)=\psi(x,y)$. The resulting distribution of spacings
for the randomized system is GUE for sufficiently high values of $\phi$.
 This distribution corresponds to
a stronger suppression of small spacings than in the GOE case,
since now $P(s) \sim s^2$ for small $s$ (Eq.\ref{wd}). Fig.
\ref{psrand.fig} shows the level statistics with and without
magnetic flux, along with the theoretical GOE and GUE curves.

In the case of the perfect tiling, one cannot apply this technique
to create a magnetic flux through the samples because of the exact
reflection symmetry of our samples. This symmetry causes the
Hamiltonian to remain in the GOE class, because it is possible to
find a real symmetric representation of H even after the
introduction of a flux tube through the sample ( i.e. if R is the
operator for reflections $({x,y}\rightarrow {y,x})$ then H is
invariant under the combined operation RT where T is the time
reversal operator). GUE statistics on the octagonal
tiling can still be obtained by including an external magnetic field
perpendicular to the plane of the tiling in the Hamiltonian.

\subsection{The spectral rigidity function of the octagonal
tiling}

 Since level repulsion is present,
wavefunctions are expected to be extended in the quasiperiodic
tiling. One can expect that a wave packet initially localized in
some region will spread out over time. This is now shown by an
analysis of the spectral rigidity. In the randomized tiling, the
spectral rigidity function crosses over from the small $E$
logarithmic law given in RMT to be $\Sigma^2(E) \sim \ln(E)$
to a power law behavior. It is found that

\bea \Sigma^2(E) \propto E^{\gamma} \eea

in a large range of energies, with $\gamma \approx 1.7$. The
exponent for quantum diffusion is therefore determined according
to eq.\ref{sigma2.eq} to be $\sigma \approx 0.85$. The perfect
tiling shows a smaller value of the exponent, $\sigma \approx
0.8$. The result for the perfect tiling is thus in good accord
with Passaro et al \cite{benza2} who found an average value 0.78
for an electron diffusing in the octagonal tiling. Our value for
the random tiling is higher than the value given by those authors
of 0.81. Both calculations conclude that, interestingly,
diffusivity is $higher$ on the disordered tiling. Disorder thus
plays a role opposite to the one it plays in a crystal, in
rendering the electron more mobile.

\begin{figure}[ht]
\begin{center}
\includegraphics[scale=0.40]{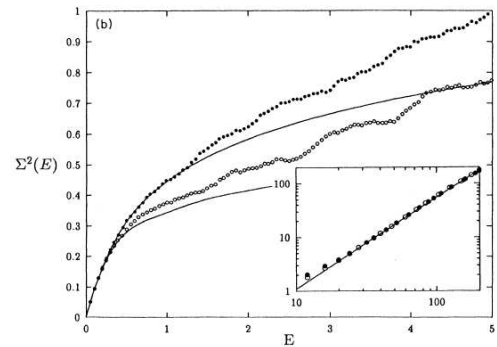}
\vspace{.2cm} \caption{Spectral rigidity of the randomized tiling (solid circles: GOE;
open circles: GUE). The curves correspond to RMT predictions. Inset: log-log plot of the power law region.
(from\cite{pj}) } \label{fred.fig}
\end{center}
\end{figure}

\subsection{Wavefunctions in the octagonal tiling}
\begin{figure}[ht]
\begin{center}
\includegraphics[scale=0.50]{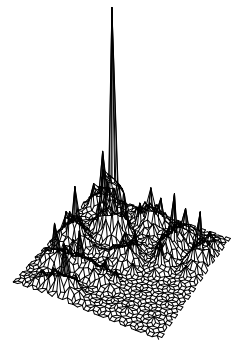}
\vspace{.2cm} \caption{Probability densities on sites of an approximant of the octagonal tiling for a state of energy $E \approx 3.98$ }
\label{wavefun2.fig}
\end{center}
\end{figure}

\begin{figure}[ht]
\begin{center}
\includegraphics[scale=0.50]{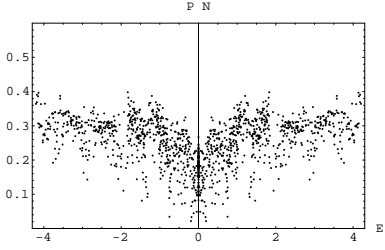}
\vspace{.2cm} \caption{Plot of p =P(Participation ratio)/N for the octagonal tiling for the approximant of 1393 sites}
\label{ipr.fig}
\end{center}
\end{figure}

Fig.\ref{wavefun2.fig} shows the probability distribution
of an electronic state near the band edge, for one of the square
approximants. One can see the typical spiky spatial dependence,
with some well-defined peaks, characteristic of critical wavefunctions.
Fig.\ref{ipr.fig} shows results for
p = P(participation ratio)/N(number of sites), plotted against the energy.  
Wavefunctions were investigated in detail on
the Penrose tiling and are reviewed in \cite{grimm}, where a similar plot is shown for the Penrose
tiling.
The data, in both the tilings, indicate that sites are more localized in the center of the
band, compared to the edges. 

\subsection{Transport in the octagonal tiling}
\begin{figure}[ht]
\begin{center}
\includegraphics[scale=0.60]{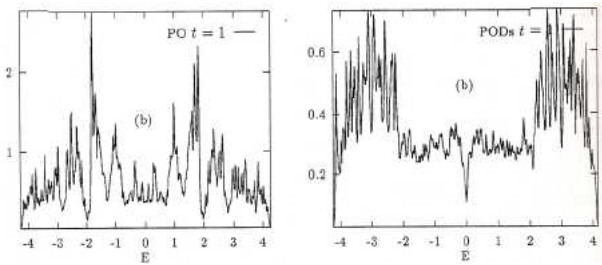}
\vspace{.2cm} \caption{Thouless number for the perfect (left) and
the randomized (right) tilings (from \cite{piechonthese})} \label{thounum.fig}
\end{center}
\end{figure}

The results of calculations of the Thouless number in the perfect
and the disordered octagonal tilings are shown in
fig.\ref{thounum.fig} (taken from
\cite{piechonthese}). The disordered tiling has, as expected,
smaller fluctuations - note the change of scale between the figures.
 The perfect tiling appears to have the
highest value of $g_{th}$ in the center of the band but the
fluctuations here are also greater. A careful study of size
dependence remains to be done for this system. Early calculations
on approximants of the  Penrose tiling \cite{tsune2} showed that
the Thouless number for that system increased towards the band
edge. However, that model is dual to the vertex type model we
consider here. Tsunetsugu and Ueda also calculated the conductance
of strips using the multichannel Landauer formalism, and found the
spiky fluctuating dependence that is a consequence of the critical
wave functions and spiky density of states. They found furthermore
that the wavefunctions evolve from power-law-extended to
power-law-localized, as energy increased -- since the tendency to
open gaps increases with the energy in their model. Zijlstra has
considered the octagonal tiling conductance by the same technique,
but includes a potential energy term in the Hamiltonian which
opens gaps in the spectrum. The results for the scaling of the
conductance with size seem to indicate that despite the gaps,
 the wavefunctions for that system decay with power laws \cite{zijlstra}.

\section{III. A family of weakly quasiperiodic tilings}

\begin{figure}[ht]
\begin{center}
\includegraphics[scale=0.60]{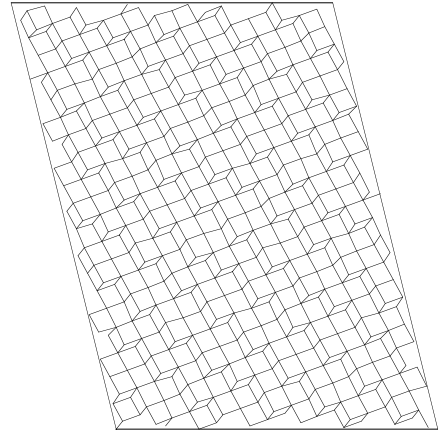}
\vspace{.2cm} \caption{Rauzy tiling - a two-dimensional projection of the cubic lattice}
\label{rauzy.fig}
\end{center}
\end{figure}

Fig.\ref{rauzy.fig} shows a structure introduced by Vidal and
Mosseri called a generalized Rauzy tiling \cite{vidmoss}(GRT). It
is obtained by projecting a cubic lattice along an irrational
direction. The sample shown is an approximant, i.e., it has been
constructed so as to allow periodic boundary conditions, but the
infinite structure is quasiperiodic. The tight binding hamiltonian
for approximants is very simple to write down contrarily to the
quasicrystals discussed earlier. This is due to the fact that this
lattice has a codimension of 1 (the codimension is the difference
$D-d$, where $d$ is the physical dimension and $D$ the dimension
of the hypercubic lattice).

For periodic approximants, the connectivity matrix is a band
diagonal (Toeplitz type) matrix. The nonzero matrix elements are
situated at distances of $F_{n-3}, F_{n-2}$ and $F_{n-1}$ from the
diagonal, where the generalized Fibonacci numbers $\{ F_{n}\}$ are
obtained from a three term recursion relation $F_{n} = F_{n-1} +
F_{n-2} + F_{n-3}$, with the initial conditions $F_{-1} = 0; F_{0}
= F_{1} =1$. The ratio $F_{n}/F_{n-1}$ tends to the value
 $\alpha \approx 1.839...$, solution of the equation
$x^3 = x^2 + x + 1$. These matrices correspond to approximants of
increasing size as $n$ is increased. For example, with $n=5$ one
has a 13 site periodic problem with a site connectivity matrix $C$
given by

\bea C = \left(
\begin{array}{rrrrrrrrrrrrr}
0&0&1&0&1&0&0&1&0&0&0&0&0 \\
0&0&0&1&0&1&0&0&1&0&0&0&0 \\
1&0&0&0&1&0&1&0&0&1&0&0&0 \\
0&1&0&0&0&1&0&1&0&0&1&0&0 \\
1&0&1&0&0&0&1&0&1&0&0&1&0 \\
0&1&0&1&0&0&0&1&0&1&0&0&1 \\
0&0&1&0&1&0&0&0&1&0&1&0&0 \\
1&0&0&1&0&1&0&0&0&1&0&1&0 \\
0&1&0&0&1&0&1&0&0&0&1&0&1 \\
0&0&1&0&0&1&0&1&0&0&0&1&0 \\
0&0&0&1&0&0&1&0&1&0&0&0&1 \\
0&0&0&0&1&0&0&1&0&1&0&0&0 \\
0&0&0&0&0&1&0&0&1&0&1&0&0 \\
\end{array}
\right) \eea

It is thus not only easy to write down the Hamiltonian for big
approximants for numerical calculations, but one has in
this indexation of the sites a convenient basis for analytical
calculations, as opposed to the octagonal and Penrose tilings.

The question now arises: are there any physical consequences of
reducing the codimension ? From the point of view of geometry it
seems clear that the fluctuations of the geometry will be reduced
when there are fewer ``degrees of freedom" in the problem. As we
saw in the preceding section, a study of the eigenvalues of the
tight binding Hamiltonian can yield information about the
wavefunctions and the dynamics of an electron in the medium. We
have already seen the use of statistical tools with which to
quantify the degree of complexity of the Hamiltonian. The results
obtained for the GRT in \cite{jagrauzy} gave a number of insights into the 
similarities and differences between this quasicrystal and the ones that had been studied before,
as we will now describe.

\begin{figure}[ht]
\begin{center}
\includegraphics[scale=0.40]{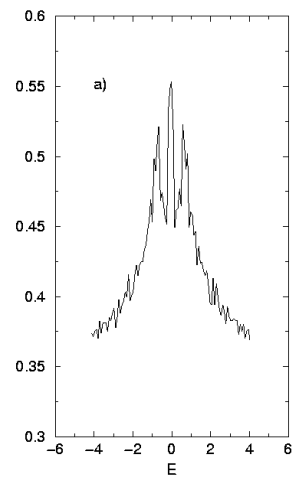}
\vspace{.2cm} \caption{Density of states of the GRT}
\label{dosrauzy.fig}
\end{center}
\end{figure}

After diagonalization of the Hamiltonian, one obtains the spectrum
and the DOS as shown in the figure \ref{dosrauzy.fig}. The DOS
rises sharply towards the band center, recalling the van Hove
singularity of the square lattice (and one sees in
fig.\ref{rauzy.fig} that the GRT has many small patches of sites
of coordination number 4). The DOS does not have the large
fluctuations displayed by the octagonal tiling, instead, there
appears to be an underlying smooth component to the GRT density of
states. This enables us to compute the level statistics analysis
using an unfolding technique similar to the one used for random
systems.

\subsubsection{Level statistics of the GRT}
The distribution of spacings is not of the Wigner-Dyson type. We
find instead a good fit to semi-Poisson statistics:

\bea P_{sP}(s) = 4 s \exp^{-2s} \eea

\begin{figure}[ht]
\begin{center}
\includegraphics[scale=0.40]{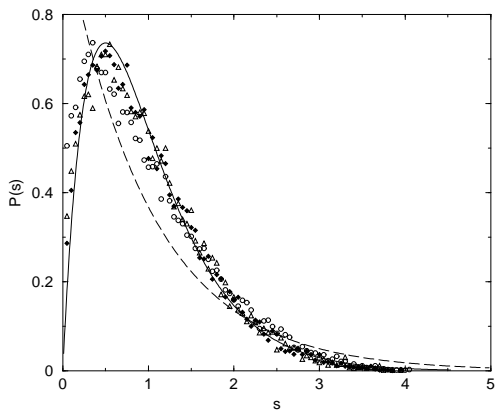}
\vspace{.2cm} \caption{P(s) of the GRT for three different sizes. Curves correspond to semi-Poisson and Poisson laws discussed in the text (from \cite{jagrauzy})}
\label{psrauzy.fig}
\end{center}
\end{figure}

Figs.\ref{psrauzy.fig} show the plot of the data obtained for the
distribution of level spacings, along
with the theoretical semi Poisson and the Poisson curves. The agreement is good over a large range of $s$,
although the data falls
off faster than the theoretical prediction at large $s$.

 In the present case of the GRT, and the results for its
$P(s)$, one can speculate that the electron motion is not
integrable as in a crystal, but not chaotic as in the octagonal
tiling. The GRT thus represents an intermediate case between the
integrable square lattice and the nonintegrable octagonal tiling.
This conclusion is in keeping with our intuition that the
quasiperiodicity is, so to say, weakened due to the small
codimension.

One expects, correspondingly, that the electron motion should
approach the ballistic limit, with the root mean square distance
travelled being linear in time $t$. This can be checked by
computing the spectral rigidity function next.

\subsubsection{Spectral rigidity of the GRT}
\begin{figure}[ht]
\begin{center}
\includegraphics[scale=0.40]{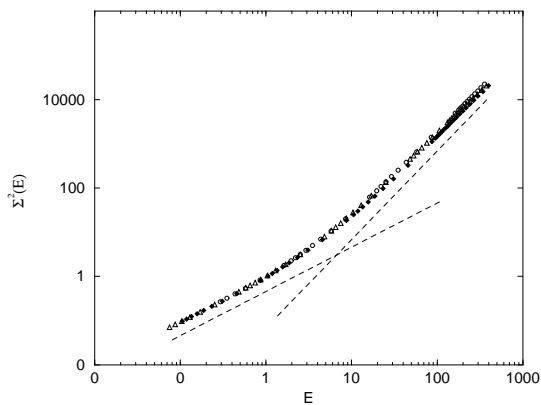}
\vspace{.2cm} \caption{Spectral rigidity of the GRT on log log scales (from \cite{jagrauzy})}
\label{sigrauzy.fig}
\end{center}
\end{figure}

The spectral rigidity corresponding to the semi Poisson case was calculated to be

\bea \Sigma^2_{sP}(E) = \frac{1}{2}[E + \frac{1}{4}(1-\exp{-4E})]
\label{sigmasp} \eea

which has a linear dependence at small energy, with a slope of
one-half. Fig.\ref{sigrauzy.fig} shows the spectral rigidity 
(without unfolding) for the GRT. The behavior found agrees for small $E$,
where it is linear, with a cross over to quadratic. The latter behavior at large $E$
(or short times) reflects the fact that the motion of the electron
is close to ballistic, as we have already surmised and the
exponent for quantum diffusion $\sigma\approx 1$.

An explicit
calculation of the root mean square distance have been carried out
by Triozon et al \cite{trio}, for wave packets of varying
energies. The effective exponent $\sigma$ found by these authors
 is about 0.9 at the center of the band, goes down to a minimum
value of about 0.85 and then increases again, approaching 1 as the
energy increases towards the band edge. The linear behavior of
$\Sigma^2$ at small energies agrees with the semi Poisson formula
given in Eq.\ref{sigmasp}.
The increased diffusivity at the band edges was found by Triozon
et al \cite{trio} to be accompanied by an increase of the
wavefunction participation ratio (PR) as energy increases. 
This is reminiscent of the behavior
 of the IPR on the
Penrose tiling described in the previous section.

One sees that the energy dependence of $\sigma$ behaves in
the opposite way of what one expects in disordered systems. There,
wavefunctions tend to be most delocalized at the band center, and
typically get more localized as one goes out to the band edge. One
can propose a handwaving explanation of this unconventional
behavior in the quasicrystal. In a disordered metal, the long
wavelength (low energies, close to the band center) modes are
relatively less scattered by the impurities, and so these
wavefunctions remain long ranged and contribute to transport. For
higher energies, the wavefunctions are more sensitive to
scattering and will tend to become more localized as the wave
vector (and the energy) increases. In the case of the
quasicrystal, for very low energies, where the wavefunction
modulation is slow, the lack of translational invariance is more
strongly felt compared to higher energies, where the wavefunction
varies on the scale of the small crystalline patches  that we
alluded to earlier. This leads one to expect that the
quasiperiodic fluctuations of geometry are $less$ effective in
localizing the wavefunction at $high$ energies than at low
energies.

\section{IV. Discussion and conclusions}

We have presented studies of the energy level statistics in
several two dimensional quasiperiodic structures and compared them
with periodic as well as disordered structures.

The strong quasicrystals such as the octagonal tiling and
also the Penrose tiling, are described by two different
forms of the level spacing distribution. There is the distribution
$P(\tilde{s})$ of the maximally unfolded spacings, where
fluctuations in the density of states have been compensated down
to the smallest energy scale, has the universal Dyson-Wigner form.
These statistics are found irrespective of whether a geometric
disorder is added or not. The tilings resemble, in this respect,
other complex systems, with classically chaotic, ergodic
trajectories, and the ``strong" quasicrystal falls into this
category. If one considers the bare unrenormalized spacings, $P(s)$ has
a log normal form for both the octagonal and the Penrose \cite{winp} tilings,
 as in some other problems involving
heirarchical processes. In this case, the log normal spacing
distribution reflects a heirarchical structure of the spectral
density, which has huge fluctuations. The DOS for the randomized
case, on the contrary, is closer to that of a typical disordered
system.

We then considered a weaker quasicrystal, the generalized Rauzy
tiling, obtained by projection of the simple cubic lattice. Here
the spacings have a different probability distribution -- the
semi-Poisson law. This tiling represents an intermediate situation
between the strongly chaotic and the integrable systems. The Rauzy
tiling is in this respect closer to the crystal than the octagonal
tiling, due to its smaller codimension.

We showed that the dynamics of the electron, as deduced from the
spectral rigidity function also follows a different behavior in
the strong and the weak quasicrystals:  propagation with an
average diffusion exponent of about 0.8 in the perfect octagonal
tiling, corresponding to a superdiffusive dynamics, while having a
value close to unity for the generalized Rauzy tiling,
corresponding to near-ballistic propagation. Disordering the perfect octagonal tiling
results in a slightly bigger value of the average exponent for anomalous
diffusion.

One can ask what occurs in the limit of increasing, instead of
decreasing, the codimension of the two dimensional tilings.
 This would result in structures
having an increasing number of local environments, and
increasingly stronger quasiperiodic disorder with $D$. The motion
of the electrons is expected to become progressively more
hampered. Destainville et al \cite{vidal2} have shown that there
is no localization even in the limit of infinite $D$. As
codimension is increased, the exponent for diffusion decreases,
and appears to tend to the value $\sigma=0.5$ i.e. ordinary
diffusion. The level spacing distribution has not been calculated,
but we would expect, based on the results for the other systems
discussed here, this to be Wigner-Dyson.

Finally, a word about the conductance properties of two-dimensional
quasicrystals. The spectral studies described here lead us to
expect that the pure hopping single component quasicrystal is close to
being a metallic
conductor. The quantum
diffusion studies show that there are huge fluctuations as a
function of the initial position of the particles, their energy,
and the time elapsed - leading to some effective average behavior
that is superdiffusive.  Similarly, huge fluctuations occur in the Thouless number
for the perfect tiling, but are attenuated in the randomized structures.
The effect on the conducting properties is not well understood.
 This effect, as
well as the detailed characteristics of this distribution remain
to be investigated in detail.

\acknowledgments We would like to thank Michel Duneau for the
fruitful discussions over many years. I would like to acknowledge as well, very
helpful discussions with I. Aleiner and B. Altshuler,
T. Ziman, J. Vidal, C. Sire and U.
Grimm. We also thank the central computing facility IDRIS on the Orsay
campus, for making these studies
possible.

\end{document}